\documentclass[12pt,a4paper]{article}
\usepackage{epsf,pifont}
\usepackage{graphicx}

\begin{document}

\baselineskip 18pt      

\begin{center}

{\bf
LIMITING TEMPERATURES IN MULTIFRAGMENTATION
}
\vspace*{0.6cm}

\normalsize                

W. Trautmann$^{(a)}$ and the ALADIN2000 Collaboration\\
\vspace*{0.3cm}
   {\small \it  $^{(a)}$ Gesellschaft f{\"u}r Schwerionenforschung (GSI),\\
Planckstr. 1, D-64291 Darmstadt, Germany }\\

\end{center}

\vspace*{0.2cm}

\begin{abstract}

The systematic data set on isotopic effects in spectator fragmentation collected 
recently at the GSI laboratory permits the investigation of the $N/Z$ dependence of 
the nuclear caloric curve which is of interest in several respects.
In particular, new light is shed on the proposed interpretation of chemical 
breakup temperatures as a manifestation of the limiting temperatures predicted by 
the Hartree-Fock model. 
The obtained results are discussed within the general context of temperature
measurements in multifragmentation reactions.

\end{abstract}

\normalsize                

\section{Introduction}
\label{sec:intro}

Limiting temperatures are a manifestation of the fragility of nuclear fragments in the
hot environment generated during energetic reactions. If the level of energy transfers   
exceeds a given limit the produced nuclei will not survive as bound objects but rather 
disintegrate into smaller 
entities, nucleons, complex particles, or very small fragments. On the other hand,
for observing fragmentation, it is necessary to provide substantial amounts
of energy in order to overcome the binding forces causing the resilience
of nuclear systems against fragment formation~\cite{tamain06,colonna97}. 
The limit up to which compound nuclei can exist will have to be exceeded. 
Multifragmentation reactions are thus characterized by a delicate energy balance.

\begin{figure}[htb]	
    \leavevmode
    \centering
    \epsfxsize=0.65\linewidth
   \epsffile{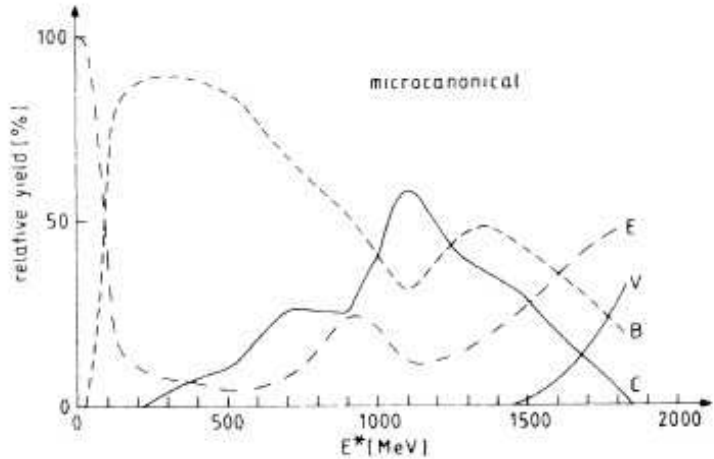}
\caption{Relative strength of the decay channels of $^{238}$U nuclei excited to energies up 
to 1800 MeV. A distinction is made between evaporation labelled E, binary decay channels 
(B, pseudo-fission), cracking (C, three or more fragments with $A \ge 20$), and 
vaporization (V, only fragments with $A < 20$). The lines give the results of a 
microcanonical Monte Carlo calculation (from Ref.~\protect\cite{gross86}). 
}
\label{gross} 
\end{figure}

The decay channels available for excited nuclei have been studied by Gross et al. 
within a phase space model \cite{gross86}. Following the principles of Weisskopf theory 
for evaporation they calculated the asymptotic phase space available for the system. The 
main assumptions in this particular model were an expanded volume in order to accomodate 
the fragments, particle emissions faster than the reaction time so that only gamma decaying 
states had to be considered, and equal a priori probabilities for the accessible states. 
The many decay channels obtained from microcanonical sampling were sorted into the classes 
of evaporation, fission, cracking, and vaporization. Cracking here denotes the appearance 
of at least three substantial fragments while vaporization indicates the complete 
disintegration into nucleons and light fragments with mass numbers $A<20$. It was found 
that cracking is the dominant reaction channel for the studied $^{238}$U nuclei at
excitation energies of the order of 1 GeV (Fig.~\ref{gross}). Cracking diminishes quickly 
at smaller and higher excitation energies which reflects the above mentioned limits. The
resulting rise and fall of fragment production predicted by this and other, similar, 
models~\cite{bond95} was found to be in good agreement with the 
experiment~\cite{ogilvie91,schuett96}.

\begin{figure}[htb]	
    \leavevmode
    \centering
    \epsfxsize=0.46\linewidth
   \epsffile{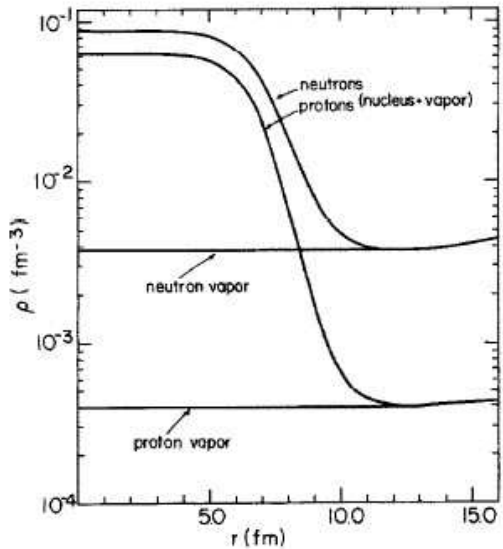}
\caption{Radial dependence of the neutron and proton densities of nucleus+vapor and vapor
solutions of the thermal Hartree-Fock equations for the uncharged $^{208}$Pb nucleus at
7 MeV temperature and with the SKM interaction in a confining spherical box with radius
$R=16$~fm (from Ref.~\protect\cite{bonche85}). 
}
\label{bonche} 
\end{figure}

\section{Limiting temperatures}
\label{sec:limit}

The temperatures to be exceeded for observing this new class of reaction processes have 
received special interest also for their connection with the nuclear equation of state. 
Calculations suggested a nearly linear relationship between the limiting temperatures at 
which compound nuclei can still exist and the critical temperature of nuclear 
matter~[7-10].
Quantitative experimental results for the 
former would thus serve as a valuable source of information on nuclear matter properties. 

The stability of hot compound nuclei 
has been studied within the temperature dependent Hartree-Fock model by several 
groups. The problem of accounting for the continuum components of the compound-nuclear
levels has been addressed by Bonche et al.~\cite{bonche85} by considering the nucleus
in thermal equilibrium with its surrounding vapor. As an example, the solution obtained 
for a $^{208}$Pb nucleus excited to 7 MeV temperature is shown in Fig.~\ref{bonche}. At 
temperatures exceeding this value, a self-bound solution is no longer found and the
nuclear matter is pressed against the boundaries of the confining volume of  
the calculation. As evident from the figure, these calculations are restricted to 
homogeneous matter distributions and spherical symmetry by the choice of the trial wave 
functions and do not explore the partition space available for fragmentation channels. 
The radii remain close to their ground-state values.

\begin{figure}[t]	
    \leavevmode
    \centering
    \epsfxsize=0.5\linewidth
    \epsffile{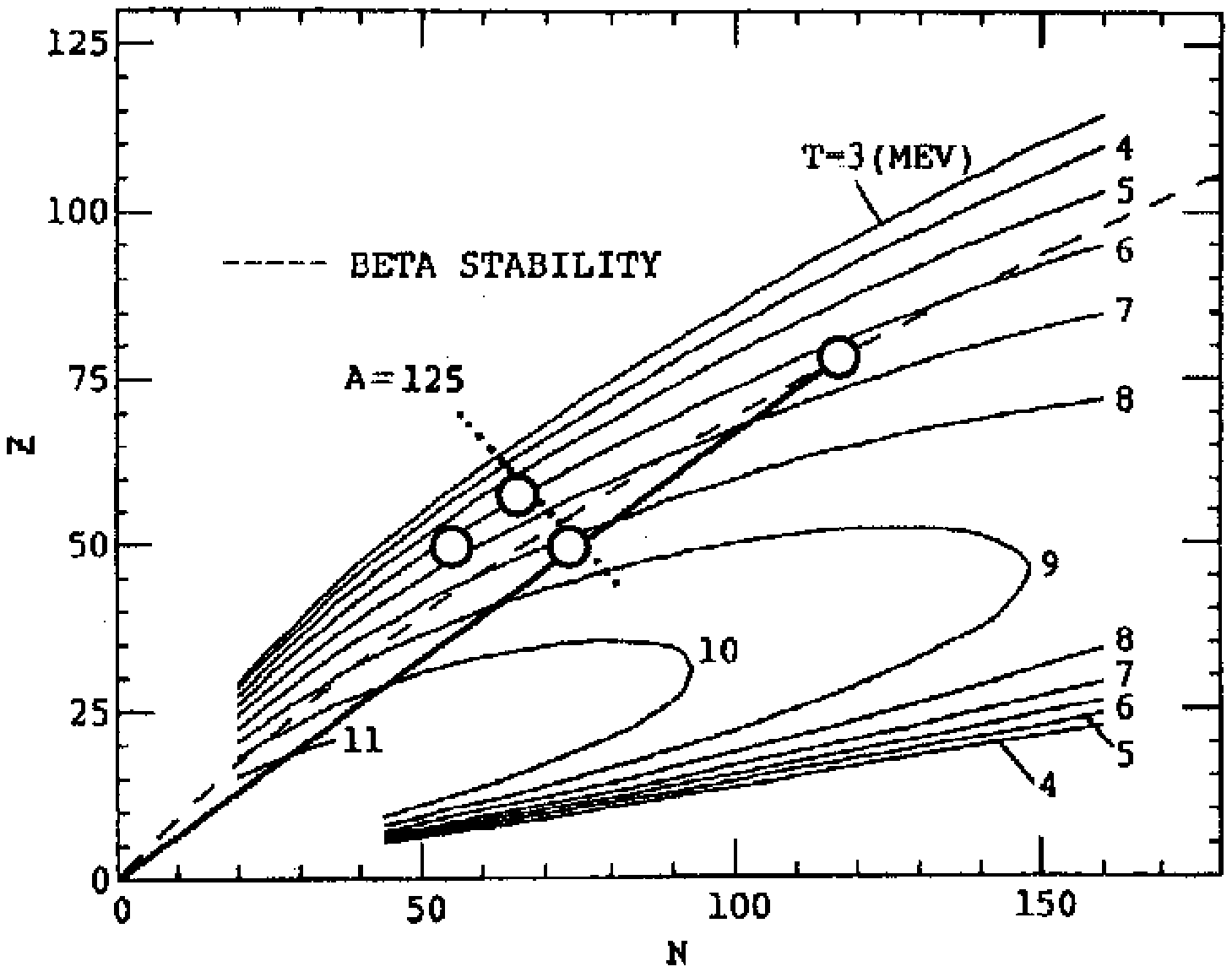}

\caption{Location of the four projectiles used in S254 
in the plane of atomic number $Z$ versus
neutron number $N$. The contour lines represent the limiting temperatures 
according to Ref. \protect\cite{besp89}, the dashed line gives the valley 
of stability, the full line corresponds to the $N/Z$ = 1.49 of 
$^{197}$Au, and the dotted line marks $A$ = 125.
}
\label{limt}
\end{figure}

A particular role in these calculations is played by the Coulomb force which if included 
drastically lowers the temperature limits. 
This is best visualized in the results of Besprosvany and 
Levit~\cite{besp89} who calculated limiting temperatures for a wide variety of nuclei 
within a liquid-drop approximation to the Hartree-Fock theory (Fig.~\ref{limt}). 
Along the valley of stability, with decreasing mass, the effect of the decreasing $Z$ 
dominates over that of the increasing $Z/A$ which permits excitations to higher   
temperatures before the system becomes unbound.
For the same reason, the limiting temperatures vary along chains of isotopes or isobars.
On the proton rich side, beyond the valley of stability, they are predicted to 
decrease rather rapidly. These variations with mass and isospin can serve as characteristic
signatures of limiting-temperature effects in experimental data.

\section{Temperature Measurements}
\label{sec:temp}

A model-free access to the temperatures governing multifragmentation reactions is 
obtained by measuring, e.g., 
excited-state populations whose interpretation relies only on the 
validity of Boltzmann statistics~\cite{poch87}. Particle-unstable resonances in light 
nuclei were found to be particularly suited for this purpose because of their stability 
against side-feeding effects during later reaction stages. 
The peak structures are identified with the technique of correlation functions, 
and background corrections are based on measurements for resonance-free pairs of 
light fragments. 

\begin{figure}[htb]	
    \leavevmode
    \centering
    \epsfxsize=0.35\linewidth
   \epsffile{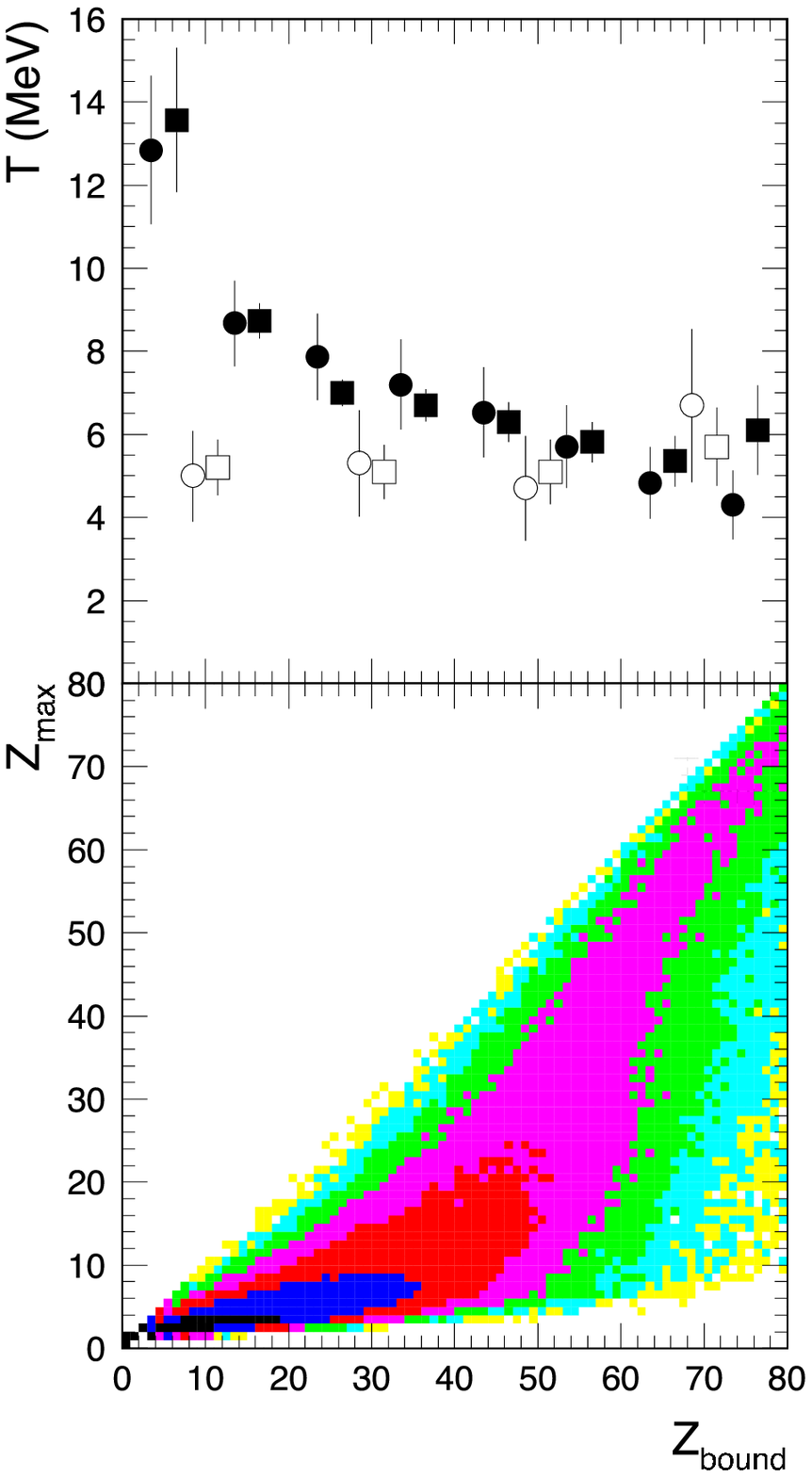}
\caption{Results obtained for $^{197}$Au on $^{197}$Au at $E/A$ = 1000 MeV:
Top: Measured isotope temperature $T_{\rm HeLi}$ (full circles and
squares representing the data for the target and projectile decays, respectively)
and excited-state temperatures (open circles and squares representing the 
data for $^5$Li and $^4$He, respectively) as a function of 
$Z_{\rm bound}$. Average values for 10 or 20 unit intervals of $Z_{\rm bound}$
are given. For clarity, the data symbols are slightly displaced horizontally.
The indicated uncertainties are mainly of systematic origin.
Bottom: Distribution of $Z_{\rm max}$ versus $Z_{\rm bound}$ after
conventional fission events have been removed. The
shadings follow a logarithmic scale (from Ref.~\protect\cite{traut07}).
}
\label{zbound} 
\end{figure}

An example of internal temperatures
measured for the fragmentation of projectile spectators from $^{197}$Au on $^{197}$Au 
reactions at 1000 MeV per nucleon is given in Fig.~\ref{zbound}. 
The open symbols represent the temperatures deduced from the ratios of 
state populations in $^4$He and $^5$Li. 
They remain at a mean value of $\approx$ 5 MeV, 
virtually independent of the variable $Z_{\rm bound}$ used for the event sorting 
according to impact parameter. 

More recently, the measurement of internal fragment temperatures has been extended to 
correlations between fragments and light particles~\cite{marie98,hudan03}. The emerging
picture is overall very consistent. For the reaction $^{124}$Xe on Sn at 50 MeV per 
nucleon, the excitation energies of fragments over a wide range of $Z$ have been found 
to saturate 
at $E/A \approx 3$~MeV which corresponds to very similar temperatures near $T=5$~MeV. 
These experimental methods probe the excitation of the fragments when they 
are sufficiently separated from the rest of the system, so that the correlations
of the decay products are not disturbed by further collisions. It is, therefore, a
question whether they represent the static limit beyond which nuclei become unstable
or rather a temporary property in a dynamical situation. A qualitative 
understanding can be obtained from the time scales of evaporation processes. 
Temperatures between 4 and 6 MeV correspond very generally to nuclear life times 
of the order of 50 to 200 fm/c \cite{borderie92,suraud06} which cover the range of 
time scales deduced for fragmentation reactions.

Access to earlier reaction stages is obtained from the isotope temperatures reflecting
the chemical freeze-out stage of the reaction. They are deduced from double ratios 
of yields of neighboring isotopes, as described previously \cite{poch95,isobook}. 
In the example of the $^{197}$Au on $^{197}$Au reaction shown in Fig.~\ref{zbound}, the 
isotope temperature $T_{{\rm HeLi}}$ rises up to about 12 MeV with 
decreasing $Z_{\rm bound}$. Its deviation from the internal temperatures starts
approximately where the drop of $Z_{\rm max}$ with respect to $Z_{\rm bound}$ indicates
the transition from the evaporation to the cracking mode of decay (Fig.~\ref{zbound},
bottom panel).
The reaction events with higher temperatures, at the smaller $Z_{\rm bound}$, are from 
sources with smaller mass, in accordance with the geometrical participant-spectator 
scenario. This implicit mass dependence has been interpreted by Natowitz et 
al.~\cite{nato95} as
suggesting that the measured isotope temperatures may reflect the limiting 
temperatures of compound nuclei for which a similar mass dependence is expected.

\section{ALADIN Experiment S254}
\label{sec:exper}

The ALADIN experiment S254, conducted in 2003 at the SIS heavy-ion synchrotron,
was designed to study isotopic effects in projectile 
fragmentation at relativistic energies. Besides stable $^{124}$Sn and $^{197}$Au beams,
neutron-poor secondary Sn and La beams were used in order 
to extend the range of isotopic compositions beyond that available with stable 
beams alone (cf. Fig.~\ref{limt}). 
The radioactive secondary beams were produced  at the fragment 
separator FRS \cite{frs92} by the fragmentation of primary $^{142}$Nd 
projectiles of about 900 MeV/nucleon in a thick beryllium  target. 
The FRS was set to select $^{124}$La and, in a second part of the experiment, 
also $^{107}$Sn projectiles which were then delivered to the ALADIN setup.
All beams had a laboratory energy of 600 MeV/nucleon and were directed onto
reaction targets consisting of $^{\rm nat}$Sn (or $^{197}$Au for the case of $^{197}$Au 
fragmentation). The acceptance of the ALADIN forward spectrometer, at this energy,
is about 90\% for projectile fragments with $Z=3$, increases with $Z$ and exceeds
95\% for $Z=6$ and heavier fragments~\cite{schuett96}.

In order to reach the necessary beam intensity of about 1000 particles/s with
the smallest possible mass-to-charge ratio $A/Z$, it was found necessary to
accept a distribution of neighboring nuclides together with the requested
$^{124}$La or $^{107}$Sn isotopes. 
While the mass and charge of each projectile can be known from measurements with
upstream detectors~\cite{luk08}, only results obtained with the full distributions 
will be presented here. 
Their mean compositions were $<$$Z$$>$ = 56.8 (49.7) 
and $<$$A/Z$$>$ = 2.19 (2.16) for the nominal $^{124}$La ($^{107}$Sn) beams, 
respectively. Model studies consistently predict that the same 
$<$$A/Z$$>$ values are also representative for the spectator systems emerging 
after the initial cascade stage of the reaction. In particular, the differences 
between the neutron-rich and neutron-poor cases are expected to remain the same 
within a few percent~\cite{botv02,lef05}.

\section{$N/Z$ dependence of the caloric curve}
\label{sec:calcurve}

The mass resolution obtained for projectile fragments entering into the 
acceptance of the ALADIN spectrometer is about 3\% for fragments with $Z \le 3$
and decreases to 1.5\% for $Z\geq 6$ (standard deviations).
Masses are thus individually resolved for fragments with atomic 
number $Z \leq 10$~\cite{bianchin_bormio}.
The elements are resolved over the full range of atomic numbers up 
to the projectile $Z$ with a resolution of $\Delta Z \leq 0.2$ obtained with the
TP-MUSIC IV detector~\cite{sfienti_prag}.

\begin{figure}[htb]     
    \leavevmode
    \centering
    \epsfxsize=0.65\linewidth
   \epsffile{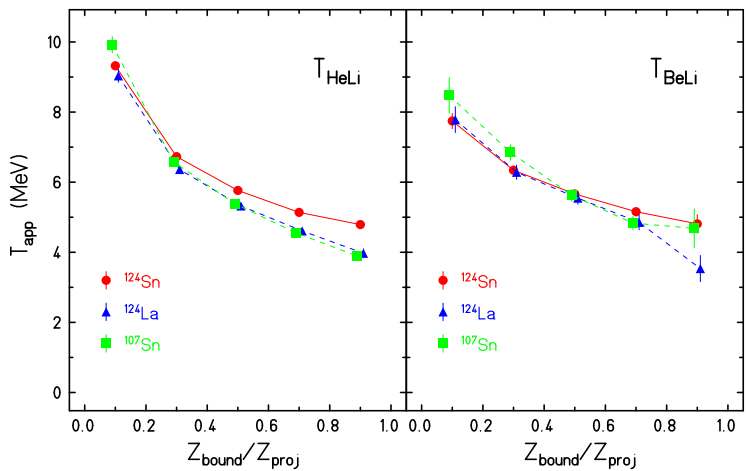}

\caption{Apparent temperatures $T_{\rm HeLi}$ (left panel) and $T_{\rm BeLi}$ 
(right panel) as a function of $Z_{\rm bound}$ for the three reaction systems
produced with $^{107,124}$Sn and $^{124}$La projectiles. 
Only statistical errors are displayed.
}
\label{temp} 
\end{figure}

Two examples of double-isotope temperatures deduced from the measured isotope yields 
are shown in Fig.~\ref{temp} as a function of $Z_{\rm bound}$. Besides the frequently
used $T_{\rm HeLi}$ (left panel), determined from $^{3,4}$He and $^{6,7}$Li yields also 
the results for $T_{\rm BeLi}$ are displayed (right panel). For $T_{\rm BeLi}$ the 
isotope pairs of $^{7,9}$Be and $^{6,8}$Li are used which each differ by two neutrons. 
The double difference of their binding energies amounts to 11.3 MeV and is nearly as 
large as the 13.3 MeV in the $T_{\rm HeLi}$ case. The apparent temperatures are displayed,
i.e. no corrections for secondary decays feeding the ground states of these nuclei 
are applied.

Both temperature observables show the same smooth rise with increasing centrality that
was observed earlier in the study of $^{197}$Au fragmentations (Fig.~\ref{zbound}). 
For $T_{\rm BeLi}$, the statistical uncertainties are larger in the two extreme bins
at large and small $Z_{\rm bound}$ in which intermediate-mass fragments are not abundantly 
produced. $T_{\rm HeLi}$, on the other hand, which mainly reflects the
behavior of the $^3$He/$^4$He yield ratio \cite{traut07} 
reaches to somewhat higher temperatures at small $Z_{\rm bound}$.
The dependence on the isotopic composition is very small and virtually non-existent in the 
fragmentation channels. At the larger $Z_{\rm bound}$, for which residue production
dominates, the temperatures of $^{124}$Sn decays are slightly larger than those
of the neutron-poor systems. This tendency is more pronounced in $T_{\rm HeLi}$ than in
$T_{\rm BeLi}$.

For a quantitative comparison with the Hartree-Fock predictions, 
the region of transition from residue production to multifragmentation 
($Z_{\rm bound}/Z_{\rm proj} \approx 0.7$) seems best suited. 
The residue channels associated with the highest temperatures are found here, 
resulting from the decay of spectator systems with about 75\% of the projectile 
mass~\cite{poch95}. Their limiting temperatures of 9.2 MeV and 7.9 MeV for the
neutron-rich and neutron-poor cases, respectively, are higher than those of the nominal 
nuclei while their difference is somewhat smaller (Fig.~\ref{limt}). 
The experimental mean values of $T_{\rm HeLi}$ and $T_{\rm BeLi}$, 
after applying a 20\% side-feeding correction~\cite{traut07,poch95}, 
are 6.2 MeV for $^{124}$Sn and 5.7 MeV for the neutron-poor systems. 

\begin{figure}[t]     
    \leavevmode
    \centering
    \epsfxsize=0.5\linewidth
    \epsffile{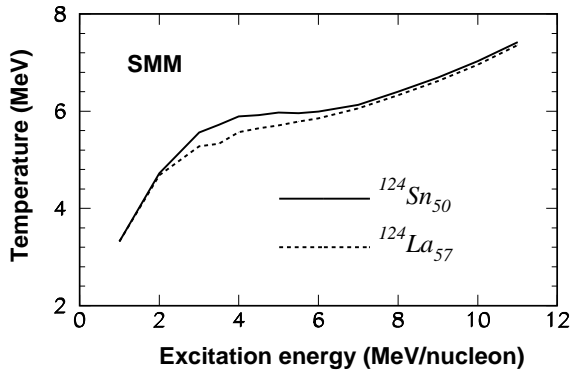}

\caption{Caloric curves for $^{124}$Sn ($Z$ = 50) and $^{124}$La ($Z$ = 57) as
calculated with the Statistical Multifragmentation Model
(from Ref. \protect\cite{ogul02}).
}
\label{calc}
\end{figure}

The large difference in absolute 
magnitude between the Hartree-Fock limiting temperatures and the side-feeding corrected 
double-isotope temperatures are not as crucial as it may appear at first sight. As noted 
already by Natowitz et al.~\cite{nato95}, the predictions depend sensitively on the type 
of Skyrme force used in the calculations~\cite{song91,kelic06}.
A linear rescaling with a factor 0.71, approximately corresponding to the results obtained
by Song and Su with the SkM$^*$ force~\cite{song91} leads to the overall best agreement in 
absolute magnitude. The corresponding critical temperature for infinite nuclear matter is
$T_c = 15$~MeV. However, even on the reduced scale, the predicted $N/Z$ dependence,
$\Delta T \approx 0.9$~MeV for the studied systems, is considerably larger than 
the experimental value. Furthermore, the interpretation of the 
$Z_{\rm bound}$ dependence as being caused by the mass dependence~\cite{nato95}
would require that the temperature difference remains on a similar level for the full range 
of $Z_{\rm bound}/Z_{\rm proj} < 0.7$ which is clearly not observed. 

A much reduced role for Coulomb effects is predicted by models which include
expansion~\cite{samad07} or which allow the partitioning of the system. 
As noted long ago by Barz et al.~\cite{barz87}, the limiting temperatures emerging from the 
Hartree-Fock approach are systematically too high because they are not going beyond
the mean-field approximation and the formation of clusters in the excited medium is not
considered. The Coulomb forces act with reduced strength in the expanded 
volume and are partly compensated by the nuclear forces within the formed fragments. 
Calculations performed more recently~\cite{ogul02} with the Statistical 
Multifragmentation Model~\cite{bond95} predict much smaller differences for the breakup 
temperatures of neutron rich and neutron poor systems with a dependence on the excitation
energy that is in excellent agreement with the experimental data (Fig.~\ref{calc}).

The open question that remains to be answered concerns the nature of the instability
driving the system into the expanded configurations sampled in the statistical
approaches. Following the statistical principles cited by Gross et al.~\cite{gross86}, 
the mere opening of the asymptotic phase space will suffice, leading to a
phase-space driven instability rather than the Coulomb instability appearing from the
Hartree-Fock simulations. Furthermore, taking recourse to recent dynamical studies, 
one observes that the partitioning of the system with a loss of binding between the
emerging parts may occur rather early in the collision~\cite{zbiri07,lefevre08}.

\section{Summary}
\label{sec:summ}

The temperatures limiting the existence of self-bound nuclei or nuclear fragments
appear in different roles in fragmentation reactions. The limits have to be overcome
to initiate the disintegration of the produced excited systems while they should
not be exceeded by the fragments to be observed. A quantitative knowledge of these
limits is not only of interest for the understanding of the reaction mechanisms 
involved but also for the more fundamental reason that they can provide information
on the nuclear-matter equation-of-state. Hartree-Fock calculations predict a linear 
relation between the limiting temperatures of excited compound nuclei and the critical 
temperature limiting the liquid-gas phase transition of nuclear matter.

The proposed interpretation of breakup temperatures measured in spectator fragmentation 
as representing the limiting temperatures was derived from a common mass dependence.
This is only partially confirmed by the new experimental data obtained from the study 
of isotopic effects in spectator fragmentation. The temperatures are smaller than the
general level of the predictions and, in particular, their variation with $N/Z$ is
smaller than expected. From the highest temperatures observed for compound channels,
a critical temperature of about 15 MeV is deduced. It should be considered as a lower 
limit because phase-space driven instabilities may initiate the disintegration 
before the static compound limit is reached. The good agreement with predictions of 
the Statistical Fragmentation Model supports this assumption.

\end{document}